# A New Method for Measuring the H I Gunn-Peterson Effect at High Redshift

Yihu Fang and Arlin P. S. Crotts

Department of Astronomy, Columbia University, New York, NY 10027

Electronic mail: fang@eureka.phys.columbia.edu    arlin@eureka.phys.columbia.edu

## ABSTRACT

As part of an undergoing program to carry out a systematic and thorough measurement of the neutral hydrogen Gunn-Peterson effect at high redshift, a quantitatively testable and repeatable procedure, particularly, a robust statistical weighting technique, is developed. It is applied to an echelle spectrum of resolution $\sim$ 15 km s$^{-1}$ of the quasar PKS 1937-101 with $z_{\rm em} = 3.787$, to demonstrate its capability.

Based on good low resolution spectrophotometric data, the continuum is extrapolated from redward of Ly$\alpha$ emission line in an objective way such that the systematic fractional deviation is minimized and estimated to be less than $\pm$ 1%. A weighted intensity distribution which is derived overwhelmingly from pixels close to the continuum level in the Ly$\alpha$ forest region is constructed by the evaluation of how closely correlated each pixel is with its neighboring pixels. The merit of such weighted distribution is its strong and narrow peak compared to the unweighted distribution, as well as its smaller dependence on uncertainty of strong absorption lines and noise spikes. It is more robust in locating the continuum level especially at higher redshift when the unweighted intensity distribution becomes flat due to rapidly increasing number of Ly$\alpha$ absorbers with redshift. By comparison to the weighted intensity distribution of synthetic Ly$\alpha$ forest spectrum with various chosen diffuse neutral hydrogen opacities $\tau_{\rm GP}$, a $\chi^2$ fit is performed between observed and model distribution. In addition to a weak line population with power law column density $N_H$ distribution $\beta = 1.7$ extrapolated down to $10^{12}$ cm$^{-2}$, a best $\chi^2$ fit requires a component of $\tau_{\rm GP} = 0.115 \pm 0.025$ at $\langle z_{\rm abs}\rangle \approx 3.4$ with estimation of the contribution from the variance of the parameter $\beta$. However, although no evidence of more than 1-2% error is seen in the continuum extrapolation, the uncertainty attributed to the possible systematic overestimation due to such an extrapolated continuum slope can be as high as the level of the accuracy of $\chi^2$ fit, which is investigated by splitting the Ly$\alpha$ forest region into subsamples to check the continuum drop's dependence on absorber redshift. Quasars at high redshift with narrow-winged emission lines are needed to improve the extrapolation and clarify any such systematic effect.

Possible contributions to the continuum drop from unresolved weak lines are discussed and it is shown in some cases the weighted intensity distribution has the potential to distinguish the continuum drop due to $\tau_{\rm GP}$ from that due to weak lines.

The future prospects of applying this technique on a larger data set and higher redshift quasars are discussed.

*Subject headings:* cosmology: observation - galaxies: intergalactic medium - quasars: absorption lines

## 1. INTRODUCTION

The Gunn-Peterson effect is the most sensitive method known for detecting smoothly distributed neutral hydrogen in the intergalactic medium (IGM). The Lyman $\alpha$ resonance scattering cross section is so large that a very small amount of diffuse neutral hydrogen in the line sight of distant quasar will dramatically depress the continuum blueward of the Lyman $\alpha$ emission line and should be easily observable (Gunn & Peterson 1965). Any null detection of such continuum drop in terms of opacity $\tau_{\rm GP}$ aside from the numerous discrete absorption lines by Lyman $\alpha$ clouds will set a striking limit for the atomic number density of diffuse neutral hydrogen in the IGM.

$$n_{\rm HI} < 2.4 \times 10^{-11} {\rm cm}^{-3}\, h\, (1+z)\, (1+2q_0 z)^{1/2}\, \tau_{\rm GP}, \qquad (1.1)$$

where $h = H_0/100$ kms$^{-1}$Mpc$^{-1}$ and $q_0$ is the cosmological deceleration parameter. It has been realized since Gunn and Peterson (1965) applied the method to quasar 3C 9 at $z_{\rm em} = 2.01$ that there is no significant trough in the continuum blueward of Ly$\alpha$ emission line. It followed immediately that with $\tau_{\rm GP} < 0.5$ at redshift of 2 the intergalactic diffuse neutral hydrogen density is $\Omega_{\rm HI} \leq 2 \times 10^{-7} h^{-1}$ which is at least 5 orders of magnitude below the baryon density inferred from the standard big bang nucleosynthesis scenario.

The apparently transparent IGM in Ly$\alpha$ imposes a powerful cosmological diagnostic to the heating and cooling process in the IGM. The universe might be highly reionized after the epoch of recombination due to the strong energy release associated with the structure formation such as quasars and primordial galaxies at high redshift. However, the studies of the thermal and ionization evolution of the IGM show that the current estimates of the number density and luminosity of quasars at $z > 3$ are insufficient to explain the absence of a high redshift Gunn-Peterson effect in terms of overlapping quasar H II regions ionizing IGM (Shapiro & Giroux 1987; Miralda-Escudé & Ostriker 1990). This suggests that either a Gunn-Peterson effect does occur for some higher redshift, the current estimate of the turn-over in quasar luminosity function for $z > 3$ is incorrect, or some unknown ionizing sources other than quasars are necessary. Although the improvement of the understanding of the quasar luminosity function at $z > 3$ can close such a discrepancy by producing a higher intergalactic ionization flux (Meiksin & Madau 1993), the earlier one observes the lack of a significant Gunn-Peterson effect, the tighter the energy production limit inferred, since more radiation per baryon is needed to explain the IGM's ionization due to the steep evolution of $\tau_{\rm GP} \propto (1+z)^{9/2}$ for constant evolution of photo-ionization radiation field (Jenkins & Ostriker 1991).

The Gunn-Peterson effect also imposes a stringent constraint on galaxy formation theories. The lack of diffuse neutral hydrogen can also be interpreted as the possibility that the baryon have already collapsed out of the IGM into gravitational bounded systems. However, the fraction of baryon from the IGM collapsed into structures depends not only on the galaxy formation scenario but also affects the evolution of the IGM. The strong reheating from galaxy formation will slow and depress further baryon collapse. The $\tau_{\rm GP}$ constraint limits such strong feed back on baryon collapse out of IGM so that substantial fraction of the total baryon density must still remain in the IGM at $z < 4$ in a Cold Dark Matter (CDM) model (Shapiro, Giroux & Babul 1994). It is seen in the numerical hydrodynamic simulation of IGM evolution that only $10^{-3}$ of the baryon have collapsed by $z = 5$ for the CDM model (Cen & Ostriker 1993).

Its extreme importance motivates us to search for such a Gunn-Peterson type of opacity in distant quasar spectra. However, previous attempts yielded only upper limits to $\tau_{\rm GP}$. In the local universe, the best limit is derived from *HST* observation of UV absorption toward 3C 273, which is $\tau_{\rm GP} < 0.1$ (Bahcall et al. 1991). At higher redshift, the most severe difficulty hampering such a measurement is caused by numerous Lyman $\alpha$ forest absorption lines blueward of the Ly$\alpha$ emission line, which make it extremely difficult to locate the local continuum while excluding contributions to the continuum drop by individual absorption lines. The general method employed is to measure the averaged fractional flux decrement between Ly$\alpha$ and Ly$\beta$ emission lines as extrapolated from continuum redward of Ly$\alpha$ emission line $D_{\rm A}$ (as defined by Oke & Korycansky (1982)). After removing the contribution to $D_{\rm A}$ from individual absorbers by calculation of the distribution of lines as functions of the rest equivalent width $W$ evolution with z, any residual of $D_{\rm A}$ is taken as a result of diffuse neutral hydrogen absorption. The previous results include a 1 $\sigma$ limit of $\tau_{\rm GP} < 0.05$ at $\langle z_{\rm abs} \rangle \approx 2.5$ from eight quasars (Steidal & Sargent 1987), and a 1 $\sigma$ upper limit $\tau_{\rm GP} < (0.11, 0.31)$ at $\langle z_{\rm abs} \rangle \approx (2.7, 3.8)$ from a compilation of quasar sample up to redshift $z > 4$ (Jenkins & Ostriker 1991).

One severe problem of such a method is that it averages out all the contributions from the absorption line systems into a single parameter that is most likely affected by the uncertainties of number density distribution as well as distribution of column density and velocity dispersion of the absorbers. Jenkins and Ostriker (1991) proposed a technique including much more information by considering how the pixel counts are distributed over intensities. The comparison of the overall shape of the distribution as well as its peak position with a similar distribution expected from a synthetic spectrum will give more robust estimations of $\tau_{\rm GP}$. (Actually, $D_{\rm A}$ only exhibits the average value of the whole distribution, so acts as a special case.) Webb et al. (1992) used this method to fit the intensity distribution of quasar Q0000-263 ($z_{\rm em} = 4.11$) with a formal detection of $\tau_{\rm GP} = 0.04 \pm 0.01$. However, since nearly all the power comes from fitting the peak of distribution, the uncertainty in determination of $\tau_{\rm GP}$ depends strongly on the width and strength of this peak. The situation deteriorates when attempts are employed on higher redshift (close to 4 or larger). The intensity distribution is entirely flattened due to the rapidly increasing number of absorbers with z in Ly$\alpha$ forest region. The power contributed to the fitting then comes increasing from other part of the distribution, which makes the result uncertain.

Aside from the difficulty to locate the local continuum in a extremely crowded Ly$\alpha$ forest region, there are two related obstacles: (1) the extrapolation of the continuum redward of Ly$\alpha$ emission line to blueward, and (2) the estimation the contributions of possible weak lines to the total continuum drop. Both have the potential uncertainties larger than the possible signal of $\tau_{GP}$. It becomes imperative to assess and minimize those uncertainties from all sources before any detection can be made.

In this paper, we propose a quantitatively *testable* and *repeatable* procedure to address aforementioned problems: an objective way to find the continuum windows redward of Ly$\alpha$ emission and to estimate the effects of *both* the fractional deviations of the extrapolated continuum level due to various choices of the continuum window (§3) and the slope which could result in overestimation of extrapolated continuum in Ly$\alpha$ forest region (§4.4), a sophisticated model assessing the contribution of weak Ly$\alpha$ lines to the continuum drop (§4.2), particularly, a powerful weighting technique to construct an intensity function for the crowded absorption spectrum that consists overwhelmingly of pixels close to the continuum level (§4.1) which in turn give us a robust measurement of the GP effect from $\chi^2$ fit (§4.3). This is the key to our program of measuring the Gunn-Peterson effect at high redshift. In collaboration with Jill Bechtold (who has collected most of the data), we are working on a fairly large sample of quasars with high resolution ($\sim$ 1 Å) spectra at redshifts around 3.0 and echelle spectra near or greater than 4. This will ensure resolution of all the absorption features down to a rest frame equivalent width cutoff $W_0$ of 0.1 Å or better, which will clear most of the uncertainty of the continuum drop due to unresolved Ly$\alpha$ forest lines in the modest resolution survey. Good low resolution spectrophotometric data are collected for each quasar to maintain a good measurement of the global continuum from Ly$\beta$ to C IV. In this paper, we implement our analysis on one particular example of the quasar PKS 1937-101 with $z_{\rm em} = 3.787$ to demonstrate our technique from extrapolating the continuum to fitting the weighted intensity distribution function with a Monte Carlo synthetic spectrum.

## 2. OBSERVATIONS AND DATA REDUCTION

The spectroscopic observations of quasar PKS 1937-101 at coordinates of $\alpha = 19^h37^m12.6^s$, $\delta = -10°9'39''$ (1950) (Wright et al. 1991) with redshift of $z_{\rm em} = 3.787$, $V = 17.0$ magnitude (Véron-Cetty & Véron 1991) were carried out at Cerro Tololo Inter-American Observatory's (CTIO) 4 meter telescope. On the nights of 1993 April 30 and May 1, echelle spectra of PKS 1937-101 were obtained by the Red Long Camera using Tek 2048 × 2048 CCD with exposure times of 3600 seconds for a total of about 6 hours, in good transparency and seeing of about 1.1″. The echelle spectrum covers the whole Ly$\alpha$ forest region from 4363 Å to 5972 Å with an average full width half maximum (FWHM) resolution of about 15 km s$^{-1}$. Standard star HR 7596 was observed as well for preliminary flux calibration.

Spectrophotometric data were obtained by the Cassegrain spectrograph with grating KPGL #2 in first order (dispersion 3.6 Å pixel$^{-1}$ and spectral resolution $\sim$ 6 Å) using the Reticon CCD (1240 $\times$ 400) on the night of 1993 April 22. The wavelength coverage is from 4825 Å to 9143 Å with an exposure time of about 25 minutes. In order to meet the spectrophotometric condition, observation were made at an air mass of less than 1.1 with seeing around 1″ and a 6″ wide slit, and the effect of light loss due to atmospheric differential refraction was minimized by rotating the position angle of spectrograph slit perpendicular to the horizon. A nearby spectrophotometry standard star LTT 7987 (Hamuy et al. 1993) was subsequently observed under the same conditions for absolute flux calibration.

The spectra were reduced with standard IRAF packages. After bias subtraction and flat-field correction, all the spectra were sky subtracted and extracted using optimally weighted variance method as described by Horne (1986). For accurate wavelength calibration of the echelle spectrum, a comparison Th-Ar lamp spectrum was obtained immediately before and after each quasar exposure and in a way similar to the object spectrum. Each order in the echelle spectrum was flux calibrated by the standard star which was also reduced in the same way as those of the quasar. The neighboring orders were superimposed and those regions with low signal-to-noise ratio at the end of each order were abandoned before all the orders from a single exposure frame were combined. The complete spectra from individual exposures were then summed to get the final result. The echelle spectrum has a signal-to-noise ratio around 10 which is wavelength dependent with higher values at the red end due to the Ly$\alpha$ emission line (see Fig. 1).

## 3. QUASAR CONTINUUM EXTRAPOLATION

Extrapolation of the quasar continuum from redward of Ly$\alpha$ emission line to the blueward is crucial in determining the continuum drop. It is generally believed that the underlying quasar continuum is a power law ranging from the UV to the IR. However, fitting such a continuum is by no means straightforward because of the existence of many emission features that make of assignment of continuum regions difficult and sometimes arbitrary. Our spectrophotometric data have sufficiently high spectral resolution ($\sim$ 6 Å) and signal-to-noise ratio to detect almost all the emission lines between 1215 Å and 1650 Å (in the rest wavelength frame) as listed in the composite spectrum from a large number of individual quasars (Boyle 1990; Francis et al. 1991). Even the weak emission features such as O I 1302 Å and C II 1355 Å are well detected. Since the fitting of the assumed power law continuum is based on regions well away from emission line wings, so called continuum "windows", the presence of these weak emission lines is very important.

We try to locate such a continuum "window" in a systematic and objective way which is less biased by the human eye. First, we identify those emission lines thought to be contaminated regions to be Ly$\alpha$ 1216 Å/N V 1240 Å, O I 1302 Å, C II 1335 Å, Si IV/O IV] 1400 Å, C IV 1549 Å, He II 1640 Å, O III] 1663 Å and N III 1750 Å. The broad Fe II emission complexes are problematic and difficult to model, however, the closest such Fe II feature sits at about 2000 Å (see Fig. 7 of Francis et al. 1991), just out of our wavelength coverage. We then assign the regions contaminated by emission lines according to the FWHM of the Ly$\alpha$ emission line and relative widths of other emission lines with respect to the Ly$\alpha$ line width. Actually, the strength of individual emission lines varies from quasar to quasar in a way difficult to model. Since we are interested mostly in fitting the underlying continuum instead of individual emission lines, we hereby do not model each of these lines but treat them in a practically simple way. To parameterize our line exclusion method, let $\Delta_j$ be the dimensionless width of individual emission lines on a logarithmic scale, that is, if width $\Delta\lambda_j$ is much less than $\lambda_j$,

$$\Delta_j \approx \log\frac{\lambda_j}{\lambda_j - \Delta\lambda_j} \approx \log\frac{\lambda_j + \Delta\lambda_j}{\lambda_j} = \log\left(1 + \frac{\Delta\lambda_j}{\lambda_j}\right), \qquad (3.1)$$

where $j$ denotes the different emission lines and $\lambda_j = (1 + z_{\rm em})\lambda_{j,\rm rest}$. Approximately, we let $\Delta_j = C_{\rm s} \times \delta_j$ where $\delta_j$ is the relative width of each emission line in units of the Ly$\alpha$ emission line FWHM so that the coefficient $C_{\rm s}$ represents the averaged absolute strength of all lines. We crudely assign larger widths to the strongest lines such as Ly$\alpha$ and C IV. Due to the strong blending of Ly$\alpha$ and N V 1240 Å, we assign $\delta$ of Ly$\alpha$ + N V as 1.25+1.2, of C IV as 1.2 and of the rest of emission lines as 0.5. The uncertainty from the measured $z_{\rm em}$ does not introduce much change in the location of the windows. The only free parameter left is $C_{\rm s}$ which determines the extent of emission line wings excluded in fitting which in turn determines the contamination free continuum window. In Fig. 2, we show the spectrophotometric data with the fitted power law and the continuum window marked as we will describe next.

The final fit of the continuum is produced in the following way: we vary the parameter $C_{\rm s}$ gradually, from $C_{\rm s}$ small enough to include substantial emission line wings to $C_{\rm s}$ large enough to barely leave pixels between emission lines, to get various outputs of the slope of the continuum. In Fig. 3, we demonstrate the behavior of the fractional deviation of the fitted continuum versus the parameter $C_{\rm s}$ for PKS 1937-101. The fractional deviations from an arbitrary reference fitted continuum are evaluated by comparing different fitted continua in the wavelength range between Ly$\alpha$ and Ly$\beta$ emission lines with the one of an intermediate value of $C_{\rm s} = 0.005$. These, hence, are measures of the deviations of fitted continuum relative to each other with different continuum windows. The 3$\sigma$ error bars in the figure originate from the goodness of the individual fits. There is a plateau in the curve where the fractional deviations of fitted continua stay within about ± 1% for a broad range of intermediate $C_{\rm s}$ values. Meanwhile, the fractional deviations fluctuate dramatically (up to 10%) for smaller or larger values of $C_{\rm s}$ due to either the contamination of emission line wings or uncertainties from less pixels included in the fit so that the noise of the spectrum dominates. In our preliminary analysis, the behavior of the fractional deviation versus $C_{\rm s}$ is similar in other quasars of our sample, except for large $C_{\rm s}$ values the deviation sometimes fall instead of rising as in this case. Suppose the existence of an underlying power law quasar continuum, with a proper value of $C_{\rm s,best}$ which corresponds to minimum contamination of emission features, the fitted slope of continuum should not vary dramatically by small changes in the adopted continuum window, equivalent to a small perturbation of the value $C_{\rm s}$ around $C_{\rm s,best}$. This argument is well justified by the fact that such a plateau exists in the figure of fractional deviations. It also provides us a way to estimate roughly the associated uncertainties of extrapolated continuum. We choose $C_{\rm s} = 0.005$ in the final fit and the related uncertainty of the fitted continuum should be around ± 1% based on the analysis above.

An alternative method is to obtain a high resolution spectrum which has the advantage of a more certain local continuum level, but the light loss due to a narrow slit and atmospheric dispersion may affect the global continuum. There have been attempts to use local minima between emission lines after smoothing out high frequency noise in echelle resolution spectra to fit the power law continuum (Giallongo, Cristiani & Trevese 1992; Giallongo, et al. 1994). We performed this local minima method on our low resolution spectrophotometry data by identifying the obvious minima between Si IV/O IV] and C IV emission lines, the one between He II 1640 Å and N III 1750 Å and two minima on both sides of O I/Si II 1304 Å. The fitted continuum decreases locally about 1-5% because of choosing local minima, but it converges with our best fit in the Ly$\alpha$ forest region. The mean fractional deviation is about 1% in the whole region of Ly$\alpha$ forest. However, the corresponding slope of this fit (in the case of PKS1937-101) would require more absorption at lower redshift end of Ly$\alpha$ forest region than at the higher redshift end, which betrays the sign of overestimation due to the slope of extrapolated continuum (see discussion in §4.4). So we think that there is no evidence of deviation of more that 1-2% in the Ly$\alpha$ forest region from our best fit of the continuum. For comparison, we also plot this alternative fit in Fig. 2.

Since the whole Ly$\alpha$ forest region is found in both our high resolution echelle spectrum and low resolution spectrophotometry data, we can convert the pixel counts seen in the high resolution Ly$\alpha$ forest spectrum by the flux we measured at the low resolution. The spectrograph's blaze function in high resolution spectra is carefully corrected by fitting a low order Chebyshev function. It allows us to directly normalize the high resolution Ly$\alpha$ forest spectrum to the extrapolated continuum from the low resolution spectrophotometry data. The strong Ly$\alpha$ and Ly$\beta$ 1026 Å/ O VI 1034 Å emission line wings also affect the determination of local continuum in the Ly$\alpha$ forest region. The same criterion used in selecting continuum window in spectrophotometry data is applied here to exclude these emission line wings. It happens that only the portion of echelle spectrum with wavelength in the rest frame between 1046 Å and 1180 Å is used in the analysis of intensity distribution in the next section.

## 4. STATISTICAL ANALYSIS OF INTENSITY DISTRIBUTION

To measure the Gunn-Peterson effect, we must have a sensitive measure to disentangle the continuum drop due to absorption from smoothly distributed neutral hydrogen and the numerous, clumpy absorption clouds. Such an effort becomes extremely difficult at high redshift since the Ly$\alpha$ forest absorption lines are so crowded, about one line per angstrom, so as to overlap, leaving virtually no well-defined continuum even with good spectral resolution. As a consequence, the intensity distribution is flattened with no prominent peak to mark the continuum level. In the following section, we use a weighting technique to select those pixels in the intensity distribution of the crowded Lyα forest spectrum that are at or near the continuum level.

### 4.1. Weighted Intensity Distribution

The purpose of the weighting technique is to select those pixels near the continuum level by weighting the value at each pixel according to the local shape of the spectrum, particularly how closely a pixel is correlated with its neighboring pixels. Pixels are treated in a statistical way judged by the structure of absorption line profile instead of by fitting individual lines, regardless of their intensities and wavelength. Two steps are taken to pursue this goal: first, we construct an empirical weighting function at each pixel to give higher weight to those pixels at the flat part of the spectrum, which most probably lie close to the top or the bottom of absorption lines. We then use a filter to exclude those pixels at the bottom of strong lines.

The intensity distribution $n(p)$ is the count of pixels with the fractional intensity $p = I/I_c$, where $I_c$ is the extrapolated continuum from redward of Lyα emission line as described in last section. In the $i$th bin, the $n(p_i)$ is the total counts of pixels with $p$ in the range of $[p_i, p_{i+1}]$ ($i = 1, ..., N_{\text{bin}}$), $N_{\text{bin}}$ is the total number of bins of the distribution.

For the $k$th pixels with intensity $p(k)$ normalized to $I_c$, the weighting function is adopted as

$$w(k) = \frac{1}{|p(k+1) - p(k)| + \epsilon} + \frac{1}{|p(k) - p(k-1)| + \epsilon} + O(2), \qquad (4.1a)$$

the secondary part $O(2)$ is to include more neighboring pixels into the weighting function, with correspondingly less weight, e.g.

$$O(2) = \frac{1}{2}\left(\frac{1}{|p(k+2) - p(k+1)| + \epsilon} + \frac{1}{|p(k-1) - p(k-2)| + \epsilon}\right), \qquad (4.1b)$$

where $\epsilon$ is a small positive number (in this case $\epsilon = 0.005$) to prevent the divergence of $w(k)$ as the $p(k)$ of neighboring pixels equal each other.

Those pixels at the strong saturated absorption line trough have similar behavior in the weighting function $w(k)$ as those at the continuum level. To exclude the effects of strong saturated lines, we try to filter out those pixels by a second derivative filter. Ideally, for a smoothly varying function such as an absorption line profile, those points at the absorption trough have negative values of the second derivative, while those at top have positive ones. Since the second derivative measure is very sensitive to local smoothness, the noisy spectrum has too many bumps to make such a measure unambiguous to its underlying true profile shape. To overcome this, at each pixel, we bin $M$ neighboring pixels and use binned pixel values to calculate the second derivative of the line profile at every point. A negative cut-off is set to exclude those pixels in troughs. We choose $M$ pixels to correspond to a few Å, which is a characteristic line width of strong lines in Ly$\alpha$ forest. Such filtering works extraordinarily well, in a statistical sense, to locate almost all of the pixels near the trough. Those trough pixels in lines of wider width will not be filtered out well, e.g., the damped Ly$\alpha$ lines can have line widths as high as a few $\times$ 10 Å which have to be removed by hand.

After the second derivative filtering and application of the weighting function $w(k)$ at each pixel $k$, we produce a weighted intensity distribution $S(p)$ with relatively more power contributed from those pixels near the continuum level. At the same $i$th bin, each pixel $p_i(k)$ will be assigned a weight corresponding to the weighting function calculated in eqs. (4.1a) and (4.1b), $w_i(k)$, so that the weighted intensity distribution will be

$$S(p_i) = f \sum_1^{n(p_i)} w_i(k), \qquad (4.2)$$

where the fraction $f = \sum_i n(p_i) / \sum_{ik} w_i(k)$ is a normalization factor to convert the weighting function to a relative weight instead of an absolute weight.

This empirical weighting function not only tends to avoid those pixels at the steep absorption profiles by assigning small weight to them, but also avoids being affected strongly by noise spikes. The neighboring pixels of noise spikes are not physically correlated and their values vary significantly from one to the other. This produces a small weight. Only those pixels with a few consecutive, slowly varying neighbor pixels are given higher weight, and thereby, dominate the weighted intensity distribution. To test the legitimacy of whether the weighting technique would have the tendency to pick up noise spikes that shift the position and distort the intensity distribution, we construct a model spectrum of a constant continuum with noise. The resulted intensity distribution of the noise spectrum is a gaussian function as expected. After applying the weighting technique, the new intensity distribution also remains symmetric around the same central peak position without any shift. This also suggests that higher signal-to-noise Ly$\alpha$ forest spectra will have better weighted intensity distributions since less pixels are underweighted as noise spikes.

In Fig. 4, we compare the weighted intensity distribution $S(p_i)$ of PKS 1937-101 with the conventional unweighted one. For the sake of direct comparison, the histogram of the distribution is normalized with unit area under the curve. The weighted intensity distribution has a striking peak which is overwhelmingly populated by those pixels close to the continuum level so that it peaks toward the direction of high intensity. The variance of that distribution is much smaller than the unweighted one (less than 2/3), which gives a narrower width. As we will see in the next two sections, when we try to compare Monte Carlo simulations to observation data, the $S(p_i)$ with a prominent peak has a great advantage in improving the $\chi^2$ fit. Any small amount of mismatch of the distribution peak will result in significant difference in the $\chi^2$ distribution. It also depends much less on the population of strong and saturated lines which are difficult to model since those lines are on the tail of the power law column density distribution and their population fluctuates strongly from quasar to quasar. $S(p_i)$ will be a much more robust statistical measure of $\tau_{\rm GP}$, which depresses the continuum globally and systematically shifts the whole intensity distribution in the $p$ axis. In the next section, we will describe the Monte Carlo simulation to produce synthetic spectra with various $\tau_{\rm GP}$, then we will test the weighted intensity distribution in $\chi^2$ fits.

### 4.2. Synthetic Lyman $\alpha$ Forest Spectrum

We have generated a code to produce synthetic Ly$\alpha$ forest absorption line spectra to simulate the observed data and will implement the same weighting technique on such synthetic spectra. The method to generate the synthetic absorption line spectrum follows the conventional methods.

The Ly$\alpha$ absorbers are randomly placed between the redshifts of quasar Ly$\alpha$ and Ly$\beta$ emission lines to produce series of overlapped absorption lines with Voigt profiles. There is no attempt to include the effect of clustering of Ly$\alpha$ clouds in the velocity space. Their random distribution pattern is supported by studies of two point correlation of Ly$\alpha$ forest clouds (cf. Sargent et al. 1980), although there is significant evidences of clustering on scales smaller than 300 kms$^{-1}$ (Webb 1987; Crotts 1989). We do not attempt to identify and fit every line in our observed spectrum in order to find the parameters to be used in our synthetic spectrum. Rather, we choose reasonable parameters from previous studies of larger samples. Our future large sample of spectra will also be treated in this uniform way.

The number density of Ly$\alpha$ forest absorption lines evolves with redshift as

$$\frac{dn}{dz} = A(1+z)^{\gamma}. \qquad (4.3)$$

From the study of a large sample of quasars, Bajtlik, Duncan and Ostriker (1988) have found $\gamma = 2.36 \pm 0.40$ for lines with rest equivalent width $W_0 \geq 0.36$ Å, while with the same threshold, Lu, Wolfe and Turnshek (1991) found $\gamma = 2.75 \pm 0.29$ for a larger sample. The compilation from echelle data has a smaller value (but larger uncertainty) of $\gamma = 1.68 \pm 0.80$ (Rauch et al. 1992). Bechtold (1994) use an equivalent width threshold $W_0 \geq 0.32$ Å for 34 high redshift quasars with moderate resolution spectra to get a slower evolution with $\gamma = 1.89 \pm 0.28$. The discrepancy in determination of the redshift evolution rate may originate mostly from the ambiguity of different line selection methods, different choices of line equivalent width thresholds and line blending corrections. It is still uncertain whether the redshift evolution of number density depends on line strength and which one evolves faster, weaker lines or stronger ones (cf. Weymann 1992). This situation is more notable when de-blending of absorption lines becomes more difficult in high redshift quasar spectra. However, by comparison of the above mentioned studies and combining their own line fitting of Q0000-263 ($z_{\rm em} = 4.11$) and Q1442+101 ($z_{\rm em} = 3.54$), Frye et al. (1993) obtain a best fitting value of $\gamma = 2.59 \pm 0.49$ with equivalent width threshold $W_0 \geq 0.21$ Å. Considering the accompanying uncertainties, they found that all of the results are acceptably fit in terms of different statistical tests. In our model, we choose $\gamma = 2.37$ and the normalization factor $A = 10.5$ for those lines with column density $\log N_H \geq 13.3$, which is converted (assuming $\gamma = 2.37$) from high resolution echelle data of Q0014+813 (Rauch, et al. 1992).

The high resolution echelle data make it possible to directly measure the column density and Doppler parameter using Voigt line fitting of individual lines. This yields a consistent power column density distribution

$$\frac{dn}{dN_H} \propto N_H^{-\beta}, \qquad (4.4)$$

with $\beta = 1.5 - 1.9$. The analysis by Rauch et al. (1992) of the high resolution echelle spectrum of Q0014+813 ($z_{\rm em} = 3.38$) gives a steep distribution of $\beta = 1.72 \pm 0.05$ for lines with $\log N_H \geq 13.3$. While another sample (Q2206-199, $z_{\rm em} = 2.56$) from a similar analysis gives a slightly flatter $\beta = 1.62 \pm 0.08$ distribution (Rauch et al. 1993). A recent study of high signal-to-noise and moderate spectral resolution data of $z_{\rm em} > 3.5$ quasars' Ly$\alpha$ forest (Frye et al. 1993) supports a higher value of $\beta \sim 1.7 - 1.9$ which is consistent with those derived from Voigt fitting of high resolution echelle data. Hence, we test three values, $\beta = 1.5, 1.7, 1.9$ in our model for lines within $13.3 \leq \log N_H \leq 17$. There is still controversy over whether there is a single power law extending to high values of column density. Since we are interested only in the measurement of a continuum drop due to many individual Ly$\alpha$ forest lines, we do not include lines with $N_H$ greater than $10^{17}$ cm$^2$ or damped Ly$\alpha$ systems in our synthetic spectra. In order to take into account the effect of unresolved weak lines, we extrapolate the same column density power law distribution down to line population of $\log N_H = 12$. More details of the weak line population are discussed in §5. We use an appropriate Doppler parameter distribution which is gaussian with mean value of $\bar{b} = 36$ km s$^{-1}$ and dispersion $\sigma_b = 16$ km s$^{-1}$ and cut off between 5 and 100 km s$^{-1}$. We do not assume any correlation in $b$ and $N_H$ distribution, as well as any redshift evolution in their distributions.

The synthetic spectrum is then convolved with a gaussian instrumental profile to get the appropriate spectral resolution. For the observed spectrum, we calculate the autocorrelation function of those sections of spectrum which are free of strong absorption lines so that we can measure the FWHM of those lines just resolved, which is close to the spectral resolution. The width of the instrumental profile being convolved is judged from measurements of the autocorrelation function to make our synthetic spectrum's spectral resolution close to that of the real data.

The noise is then added to the convolved synthetic spectrum. After scaling the normalized synthetic spectrum to real counts $I(\lambda)_{\rm in}$, with chosen different background levels $B(\lambda)$, we can add Poisson noise to get final spectrum $I(\lambda)_{\rm out}$ with different signal-to-noise ratios.

### 4.3. $\chi^2$ Fit and Gunn-Peterson Effect

Once we have the way to simulate the Ly$\alpha$ forest spectrum, we can add the diffuse neutral hydrogen opacity $\tau_{\rm GP}$ in the model absorption opacity. We generate a series of synthetic Ly$\alpha$ forest spectra of a quasar at redshift $z_{\rm em} = 3.787$ with its absorption lines' physical parameters chosen as those described in §4.2. These spectra have the same spectral resolution as observed data of PKS 1937-101 and similar signal-to-noise ratio $S/N = 10$, except opacity $\tau_{\rm GP}$ is chosen variously as 0.05, 0.075, 0.1, 0.115, 0.125, 0.15, 0.2. In Fig. 5(a), we plot the weighted intensity distribution of synthetic spectrum with various $\tau_{\rm GP}$. The weighted intensity distribution of observed spectrum of PKS 1937-101 is also plotted against each model for comparison (dashed line). In Fig. (5b), the same plots are shown the unweighted intensity distribution.

Given the intensity distribution of the observed quasar Ly$\alpha$ forest spectra $n_q(p)$ and $S_q(p)$ as well as a set of intensity distributions $n_s(p,\tau)$ and $S_s(p,\tau)$ of the corresponding synthetic spectrum with different $\tau_{\rm GP}$, we can distinguish statistically between observed data and model data to check whether $n_q(p)$ and $n_s(p,\tau)$ or $S_q(p)$ and $S_s(p,\tau)$ are draw from the same population distribution function. The $\chi^2$ test in the case of two binned data sets with unequal number of data points (Press et al. 1992) is,

$$\chi^2(\tau) = \sum_i \frac{(\sqrt{n_q/n_s}\, n_q(p_i) - \sqrt{n_s/n_q}\, n_s(p_i,\tau))^2}{n_q(p_i) + n_s(p_i,\tau)}, \qquad (4.6)$$

where

$$n_q \equiv \sum_i n_q(p_i); \quad n_s \equiv \sum_i n_s(p_i,\tau). \qquad (4.7)$$

The same expression also applies to the weighted intensity distribution $S_q(p_i)$ and $S_s(p_i,\tau)$.

We then get two nonlinear $\chi^2$ distributions with respect to model parameter $\tau$ for the weighted and unweighted intensity distributions. The maximum likelihood estimate of the best-fit model parameter $\tau$ is obtained by reaching the minimum point of the $\chi^2$ distribution. The general goodness of a $\chi^2$ fit is characterized by the value of $\chi^2/\nu$, where $\nu$ is the number of degrees of freedom which is the number of bins $N_B$ with non-zero data. In both our intensity distributions, we bin the data points in 25 bins covering $p = I/I_c$ from $-0.06$ to 1.38. Ideally, for a best fit, $\chi^2/\nu$ should reach a mean value of unity of a normal distribution (for large $\nu$) with standard deviation of $\sqrt{2/\nu}$.

In Fig. 6, we show the $\chi^2$ distributions derived from both weighted and unweighted intensity distributions. For the convenience of direct comparison of the merit, we plot the $\chi^2/\nu$ distribution. Overall, the $\chi^2/\nu$ distribution derived from the weighted intensity distribution $S(p_i)$ has a much better fit. The trough of $\chi^2/\nu$ distribution reaches down to around 2 for $S(p_i)$ while the corresponding minimum values are around 10 for unweighted $n(p_i)$. The improvement of the fit by a factor of 5 comes mostly from the narrowed peak of $S(p_i)$ in which the dominant contributions to $\chi^2$ are from a few bins of $p$ near the peak which is supposed to be the mark of the continuum level. Any mismatch of the peak will significantly contribute to the $\chi^2$ value and in turn the change of $\chi^2$ will robustly reflect the effect of $\tau_{\rm GP}$. In the case of the unweighted distribution $n(p_i)$, the flattened histogram does not have a prominent peak and significant contributions to $\chi^2$ can come from any $p$ position of mismatch between the two histogram. Many of these mismatches do not come from the bins near the peak, but from the tails of the distribution which include either pixels of noise spikes at the high intensity end or pixels in the trough of the absorption lines at low intensity end. This information is not directly relevant to the position of the local continuum in $p$ and depends strongly on the model parameters such as the number of strong lines per unit wavelength. So the changes of $\chi^2$ value with respect to $\tau_{\rm GP}$ from unweighted $n(p_i)$ may not necessarily reflect the direct consequence of continuum depression of various amounts $\tau_{\rm GP}$. The judgment from the fitting whether $n_s(p_i)$ from a synthetic spectrum of a particular $\tau_{\rm GP}$ matches or mismatches the observed $n_q(p_i)$ could mislead the result. This situation becomes much more severe for higher redshift ($z_{\rm em} > 4$) quasar Ly$\alpha$ forest spectra since less pixels left at the continuum produces an intensity distribution almost flat without a prominent peak and also the statistical properties of the population of lines and redshift evolution are not well studied and less certain.

We can also see that the gradient of the $\chi^2$ distribution toward the minimum point $\chi^2_{\rm min}$ is steeper in the weighted case than when unweighted. The variance at each bin of such a $\chi^2$ distribution will be $\sigma_i^2 = n_q(p_i) + n_s(p_i, \tau)$ so that we can estimate the average uncertainties associated with each of the $\chi^2$ value as

$$\bar{\sigma} = \sqrt{\sum_i \sigma_i^2 / \nu} \; . \tag{4.8}$$

The calculated 1 $\sigma$ level error bars are shown in Fig. 6. It is noticed that the typical error bars of $\chi^2$ value derived from $S(p_i)$ are only about half of those derived from $n(p_i)$. This means that the $\chi^2$ distribution derived from $S(p_i)$ is more sensitive to the parameter $\tau$.

As we can see from the Fig. 6, even though we have included an unresolved weak line population in our model line opacity with the same $N_H$ power law distribution as of the strong line $\beta = 1.7$ down to the limit $\log N_H = 12.0$ , we still cannot match the observed distribution in intensity. A diffuse neutral hydrogen opacity $\tau_{\rm GP} = 0.115 \pm 0.01$ is needed from the best fit in $\chi^2/\nu$ distribution derived from the weighted intensity distribution.

Although we have not performed a full test at this stage for the dependence of such a detection on various Ly$\alpha$ forest model parameters, we have estimated the variance of $\tau_{\rm GP}$ with the parameter $\beta$. With a power law column density distribution extrapolated down to $10^{12}$ cm$^{-2}$, the steeper distribution with a larger $\beta$ will account for more continuum drop because of more weak lines included. The observed value of $\beta$ is believed in the range of 1.5 to 1.9 (see discussion in §4.2). We generate two sets of synthetic spectra with $\beta = 1.5$ and 1.9 for various $\tau_{\rm GP}$. The same procedure is applied to get a $\chi^2$ fit between the intensity distributions of the observed and the synthetic spectra. The $\tau_{\rm GP}$ of different $\beta$ is measured from the $\chi^2$ fit of the weighted intensity distribution, which is $\tau_{\rm GP} = 0.150 \pm 0.015$ for $\beta = 1.5$ and $\tau_{\rm GP} = 0.025 \pm 0.015$ for $\beta = 1.9$. The goodness of the fit is similar to, although a little worse than, that of the $\beta = 1.7$. However, we find that the shape of the *unweighted* intensity distributions of $\beta = 1.5$ and 1.9 has noticeable deviations from the observed one, particularly at the low intensity end, where $\beta = 1.5$ gives a flatter tail and $\beta = 1.9$ has a steeper one. Since $\beta$ determines the proportion of the strong lines to the weak lines, this can be understood that a smaller $\beta$ produces relatively more strong lines so that more counts locate at the low intensity end. We use this fact to estimate the formal standard deviation in $\beta$ from the $\chi^2$ fit of the unweighted intensity distributions for fixed values of $\tau_{\rm GP}$. The $\sigma_\beta$ is about 0.04 with a mean value close to 1.67 for a wide range of $\tau_{\rm GP}$. So the contribution to the error of $\tau_{\rm GP}$ from the variance of $\beta$ is about 0.013. Considering that there is no evidence of errors larger than 1-2% from our continuum extrapolation from redward of the Ly$\alpha$ emission line, we estimate that $\tau_{\rm GP} = 0.115 \pm 0.025$. A data base larger than a single line of sight is necessary for a systematic investigation of such a test.

### 4.4. Effect of Overestimation due to Extrapolated Continuum Slope

Although the uncertainty due to the selection of continuum "windows" in our extrapolation of the continuum from redward of Ly$\alpha$ emission is minimized and estimated to be less than 2% (see §3), there is still a possible systematic effect of overestimation of the extrapolated continuum slope since the regions at the ends of the linear fit are strongly affected by the broad wings of the Ly$\alpha$ and C IV emission lines. The overestimation of the power law continuum will underestimate much more of the normalized flux $p$ at the far end of the Ly$\alpha$ forest region (closer to the Ly$\beta$ emission line) than of those at near side (closer to the Ly$\alpha$ emission line), so that it will appear to have more $\tau_{\rm GP}$ at lower redshift. We can estimate the amount of overestimation of the extrapolation by splitting the Ly$\alpha$ forest region into two subsamples: (a) 5000 Å to 5300 Å for lower redshift absorbers and (b) 5250 Å to 5550 Å for higher redshift ones. The weighted intensity distributions are produced from both subsamples and can be compared to the distribution of synthetic spectra to see how the inferred $\tau_{\rm GP}$ changes with redshift. If it increases with $z$ (or stays roughly the same with $z$ since the change of $\tau_{\rm GP}$ due to the redshift difference in this range may not be significant enough to be detected), it behaves like $\tau_{\rm GP}$ should. However, if it *decreases* significantly with $z$, it indicates that the extrapolated continuum is probably overestimated. In Fig. 7(a), we illustrate the weighted intensity distributions of these two subsamples. It is shown that both distributions peak at roughly the same position of $p$. The best $\chi^2$ fits of these distributions with synthetic spectra indicate a similar value of $\tau_{\rm GP}$ for both lower and higher redshift subsamples (Fig. 7(b)). This suggests that the possible effect of the overestimation of the continuum extrapolation are still close to the errors from the $\chi^2$ fit and the $\tau_{\rm GP}$ is unlikely to be entirely due to the overestimation. However, it can be seen from Fig 7(a) that the distribution $S(p)$ of lower redshift subsample (solid line) has some excessive power at the low intensity end. This leads to the $\chi^2$ distribution of lower redshift subsample (solid line in Fig 7(b)) flattened toward larger $\tau_{\rm GP}$ which makes the determination of $\chi^2_{\rm min}$ ambiguous. The signs of slightly more continuum drop in the lower redshift subsample suggests that at the far end of the Ly$\alpha$ forest region the normalized flux could be underestimated due to some amount of overestimation of the extrapolated continuum. Although our weighting technique has shown ability to estimate the continuum level of a crowded Ly$\alpha$ forest spectrum to an accuracy about 1-2%, in order to detect the continuum drop signal $\tau_{\rm GP}$ at a level around 0.1, we must understand the systematic effects of the extrapolated continuum within the same accuracy level. Since we do not yet have clear evidences of the evolution of $\tau_{\rm GP}$ with $z$, except theoretical arguments, any counter-evolution of the continuum drop due to overestimation will make the measurement more complicated. To ultimately clarify such a possible systematic effect, we have to look for high redshift quasars with very narrow-winged emission lines to leave longer stretches of continuum "windows" less contaminated by emission features.

## 5. DISCUSSION AND SUMMARY

One loophole still exists, which is that current observational limits on resolving the Ly$\alpha$ forest absorption lines to a complete limit is no better than a column density $\log N_H = 13.0$, even in the best echelle spectral resolution. Lines with smaller column density will be undetected but still contribute to the continuum drop. The statistical property of their distribution is unknown although we can speculate by extrapolation from a stronger line population as we did in our analysis. A conservative estimate from such a detection should be taken as combining effect from diffuse neutral hydrogen and possibly some unresolved small Ly$\alpha$ clouds unaccounted for.

Estimation of the contribution from such weak lines is extremely difficult. We manage some insight into this problem by studying the difference of the collective effect in weighted intensity distribution between a significant amount of weak line population and a component of diffuse neutral hydrogen. The collective effects of weak absorption lines will make the spectrum appear to fluctuate more, especially at the flat part which is free of strong absorption lines and supposed to be the local continuum level, while the diffuse neutral hydrogen will make a more uniform depression of the continuum. The weighted intensity distribution is much more sensitive to the fluctuation of pixel values in the flat region. To demonstrate the potential of its ability to disentangle these two effects on the continuum drop, we compare two model spectra with Ly$\alpha$ emission line at redshift 3.4: (1) with the weak line population of $N_H$ distribution $\beta = 1.7$ extrapolated down to $\log N_H = 12.0$ but without a GP opacity $\tau_{\rm GP} = 0.0$, (2) without the weak line population (with $N_H$ cutoff at $10^{-13}$ cm$^{-2}$) but with $\tau_{\rm GP} = 0.075$. The other parameters are the same in both model as those in §4.2. In Fig. 8, we illustrate the difference of these two model in the weighted intensity distribution (lower panel), for comparison, the unweighted distribution of these models are also displayed (upper panel). Two kinds of spectral resolution are simulated with higher echelle resolution about $\sim 15$ km/s in the left panel and lower resolution $\sim 75$ km/s ($\sim 1$ Å) at right. All of them have signal-to-noise ratio $S/N = 10$. We can see the remarkable difference of these two models in the echelle resolution case from the weighted distribution (left lower panel) in which the model without a weak line population appears to be more peaked, while the unweighted models are nearly identical especially at 1 Å resolution. Low resolution data will smear any fluctuation of pixel values due to weak lines and unweighted distributions may not have enough sensitivity to pick up the difference. It is encouraging that weighted intensity distribution has the potential to put constraints on the contributions from weak lines. The reason to choose a model quasar at $z_{em} = 3.4$ instead of 3.8 like PKS 1937-101 is because a higher redshift quasar has less stretch of continuum free of strong absorption left in the Ly$\alpha$ forest, so that the effect is less remarkable. It is worth studying at both lower redshift and higher ones in comparing a larger pool of spectra.

Such an effort is being undertaken on a fairly large sample of quasar Ly$\alpha$ forest spectra with redshift ranging from 2.7 to larger than 4.0 (Fang, Bechtold & Crotts 1994; Fang, et al. 1994 in preparation). Both high resolution (1 Å) spectra which are capable of detection lines to $W_0 = 0.1$ Å with 4 $\sigma$ detection limits for $z_{em} < 3.5$, or echelle spectra for higher redshift, and good low resolution spectrophotometric data are collected. The data set is split into a few groups with several quasars at roughly the same redshift range. With multiple line of sight sampling at different epochs of early universe, we can check the consistency of our whole process of the weighting technique analysis described in this paper within the same redshift group and study the possible redshift evolution of any detection of a continuum drop. Any possible systematic overestimation in the extrapolation of the continuum should apply to all quasars regardless of redshift, unless the spectral energy distribution of quasars is considerably different with redshift. Also, we can test whether $\tau_{GP}$ at a given redshift is determined to be the same regardless of the quasar redshift. Better understanding of the detected continuum drop dependence on model parameters can be investigated extensively. The contribution of a weak line population can be studied at lower redshift and then a reasonable model can be applied to higher redshift to get a tighter limit on any detection. It will be interesting to see after the systematic and thorough study whether the continuum drop we should detect follows the same power law redshift evolution as that seen for the Ly$\alpha$ forest number density. Any large deviation from this behavior may be interpreted as a true Gunn-Peterson effect.

The implementation of the weighting technique to select pixels close to the continuum level in a crowded absorption spectrum can also be applied to other spectroscopic studies, such as the high spectral resolution stellar spectrum.

To summarize the main results of this study: a quantitatively testable and repeatable procedure, particularly, a robust statistical weighting technique, has been developed and applied to an echelle spectrum of high resolution of the quasar PKS 1937-101 with $z_{em} = 3.787$. Based on good low resolution spectrophotometric data, the continuum is extrapolated from redward of the Ly$\alpha$ emission line in an objective way such that the systematic fractional deviation is minimized and estimated to be less than $\pm$ 1%. A weighted intensity distribution which is derived overwhelmingly from pixels close to the continuum level in the Ly$\alpha$ forest region is constructed by the evaluation of how closely correlated each pixel is with its neighboring pixels. The merit of such weighted distribution is its strong and narrow peak compared to the unweighted distribution, as well as its smaller dependence on uncertainty of strong absorption lines and noise spikes. It is more robust in locating the continuum level especially at higher redshift where the unweighted intensity distribution becomes flat due to rapidly increasing number of Ly$\alpha$ absorbers with redshift. By comparison to the weighted intensity distribution of synthetic Ly$\alpha$ forest spectrum with various chosen diffuse neutral hydrogen opacities $\tau_{\rm GP}$, a $\chi^2$ fit is performed between observed and model distribution. In addition to a weak line population with power law column density $N_H$ distribution $\beta = 1.7$ extrapolated down to $10^{12}$ cm$^{-2}$, a best $\chi^2$ fit requires a component of $\tau_{\rm GP} = 0.115 \pm 0.025$ at $\langle z_{\rm abs} \rangle \approx 3.4$ with estimation of the contribution from the variance of the parameter $\beta$. However, although no evidence of more than 1-2% error is seen in the continuum extrapolation, the uncertainty attributed to the possible systematic overestimation due to the extrapolated continuum slope can be as high as the level of the accuracy of $\chi^2$ fit, which is investigated by splitting the Ly$\alpha$ forest region into subsamples to check the continuum drop's dependence on absorber redshift. Quasars at high redshift with narrow-winged emission lines are needed to improve the extrapolation and clarify any such systematic effects.

We would like to thank Steve Heathcote for help observing and for collecting some of these data and Mira Véron-Cetty for supplying the digital version of their quasar catalog. This study is supported by National Science Foundation grant AST 91-16390 and the David and Lucile Packard Foundation fellowship.

– 18 –

**FIGURE LEGENDS**

Figure 1: Echelle spectrum of the quasar PKS 1937-101 with $z_{\rm em} = 3.787$ in the wavelength range between Ly$\beta$ and Ly$\alpha$ emission lines. Extrapolated continuum from redward of the Ly$\alpha$ emission line is plotted over the region of Ly$\alpha$ forest spectrum used in weighted intensity distribution analysis. Flux is in the unit of $1 \times 10^{-16}$ ergs s$^{-1}$ cm$^{-2}$ Å$^{-1}$.

Figure 2: Spectrophotometric spectrum of the quasar PKS1937-101 with fitted power law continuum (in solid straight line). The continuum windows between emission lines chosen in fitting (see §3.) are marked under the spectrum (in solid line). Square symbols marked on the spectrum

---

This preprint was prepared with the AAS LaTeX macros v3.0.

are rejected pixels in fitting with 3 $\sigma$ limits. The fitted power law with the method of choosing the local minima between the emission lines is plotted in dotted lines as well as the corresponding continuum windows.

Figure 3: The fractional deviation of the fitted power law continuum versus the $C_s$ value which is the amount of emission line excluded in fitting. The uncertainties of the extrapolated continuum is estimated within $\pm$ 1% around the adopted $C_{s,best}$ (in dashed line) for a broad range of $C_s$ values. The error bars (3 $\sigma$) are from the individual power law fit.

Figure 4: The weighted intensity distribution of the quasar PKS 1937-101 Ly$\alpha$ forest spectrum (in solid line) and the unweighted distribution (in dashed line). Here $p$ is the continuum normalized by extrapolated continuum $I_c$ from redward of the Ly$\alpha$ emission line. The distributions are normalized with unit area under the curve.

Figure 5: (a) The comparison of weighted intensity distributions $S(p_i)$ of the synthetic spectra with various diffuse neutral hydrogen opacities $\tau_{GP}$ = 0.075, 0.1, 0.115, 0.125 (in solid line) and a power law $N_H$ distribution of $\beta$ = 1.7 extrapolated down to $10^{12}$ cm$^{-2}$ to those of the observed PKS 1937-101 (in dotted line). The distributions are normalized with unit area under the curve. (b) The same plot for comparison of the corresponding unweighted intensity distribution $n(p_i)$.

Figure 6: The distribution of $\chi^2$ per degrees of freedom versus the parameter $\tau_{GP}$ derived both from the weighted intensity distribution (circle points) and the unweighted intensity distribution (triangle points) by comparison to the synthetic spectra (as noted in Fig. 5). 1 $\sigma$ error bars of $\chi^2$ values are also shown.

Figure 7: Estimation of the effect of possible overestimation of the extrapolated continuum: (a) Comparison of the weighted intensity distribution of two subsamples in the Ly$\alpha$ forest region of PKS 1937-101: lower redshift sample of 5000 - 5300 Å (solid line) and higher redshift sample of 5250 - 5550 Å (dotted line), (b) The distribution of $\chi^2/\nu$ versus $\tau_{GP}$ derived from weighted intensity distribution for these two subsamples by comparison to synthetic spectra with various $\tau_{GP}$.

Figure 8: Estimation of weak line contributions to the continuum drop from synthetic spectra of two models: weak line population of power law distribution of $N_H$ down to $10^{12}$ cm$^{-2}$ with $\beta$ = 1.7 but $\tau_{GP}$ = 0.0 (in dotted line) and no weak line population but $\tau_{GP}$ = 0.075 (in solid line). Lower panels are comparisons of the weighted intensity distribution and upper panels of the unweighted one. Left panels are from spectra of echelle resolution of 15 km s$^{-1}$ and right panels are from a lower resolution of 75 km s$^{-1}$. All spectra have S/N=10.